\documentclass[aip,jcp,notitlepage,10pt,twocolumn,superscriptaddress]{revtex4-1}
\usepackage[utf8]{inputenc}
\usepackage{geometry}
\usepackage{xcolor}

\usepackage{hyperref}
\usepackage{graphicx}
\usepackage{amsmath}
\usepackage{wasysym}
\usepackage{stmaryrd}
\usepackage{subfigure}
\usepackage{listings}
\usepackage{color}
\usepackage[T1]{fontenc}
\usepackage{tabularx}
\usepackage{multirow}

\begin{document}
\title{Avoiding critical slowdown in models with SALR interactions}
\author{Mingyuan Zheng}
\affiliation{Department of Chemistry, Duke University, Durham, North Carolina 27708, United States}
\author{Marco Tarzia}
\affiliation{LPTMC, CNRS-UMR 7600, Sorbonne Universit\'e, 4 Place Jussieu, F-75005 Paris, France}
\affiliation{Institut Universitaire de France, 1 rue Descartes, 75231 Paris Cedex 05, France}
\author{Patrick Charbonneau}
\email[Author to whom correspondence should be addressed. Electronic mail: ]{patrick.charbonneau@duke.edu}
\affiliation{Department of Chemistry, Duke University, Durham, North Carolina 27708, United States}
\affiliation{Department of Physics, Duke University, Durham, North Carolina 27708, United States}
\date{\today}
\begin{abstract}
In systems with frustration, the critical slowdown of the dynamics severely impedes the numerical study of phase transitions for even the simplest of lattice models. In order to help sidestep the gelation-like sluggishness, a clearer understanding of the underlying physics is needed. Here, we first obtain generic insights into that phenomenon by studying one-dimensional and Bethe lattice versions of a schematic frustrated model, the axial next-nearest neighbor Ising (ANNNI) model. Based on these findings, we formulate two cluster algorithms that speed up the simulations of the ANNNI model on a 2D square lattice. Although these schemes do not avoid the critical slowdown, speed-ups of factors up to 40 are achieved in some regimes.
\end{abstract}
\maketitle

\paragraph*{Introduction --} Various Monte Carlo (MC) schemes have been developed to complement local Metropolis updates in the computational study of statistical physics models \cite{newman1999monte,frenkel2001understanding,landau2021guide}.
Such enhanced configurational sampling typically relies on sidestepping energetic or entropic barriers in ways not possible with local updates alone. Designing appropriate algorithms often relies on identifying non-trivial structural features in the regimes of interest. Efficient approaches have hence been tailored for different contexts, such as cluster algorithms to displace self-assembled aggregates \cite{niedermayer1988general,edwards1988generalization,machta1995invaded,whitelam2007avoiding,ruuvzivcka2014collective}, non-local configuration-biased methods to escape traps 
created by strong inter-particle attraction \cite{chen2000novel}, and event-chain algorithms to transport defects efficiently  \cite{michel2014generalized,krauth2021event}.

Sampling configurations around simple second-order phase transitions is nevertheless difficult, given the emergence of a gelation-like percolation of the structural correlation in the critical regime. In some cases, a particularly efficient class of cluster schemes lessen the associated \textit{critical slowdown}. For the simple Ising model, for instance, the geometric construction introduced by Kasteleyn and Fortuin (KF) \cite{kasteleyn1969phase,fortuin1972random} and further expanded by Coniglio and Klein (CK) \cite{coniglio1980clusters} to accurately capture spin-spin correlations, $\langle s_i s_j \rangle$, is leveraged by the Swendsen-Wang (SW) and Wolff algorithms to markedly enhance sampling efficiency in the vicinity of the critical point\cite{swendsen1987nonuniversal,wolff1989collective,edwards1988generalization}. Straightforward generalizations of SW and Wolff algorithms, however, fail to enhance critical sampling for models with frustrated interactions, such as those used to study spin glasses and systems with competing short-range attractive and long-range repulsive (SALR) interactions \cite{binder1986spin,langer2014theories,stradner2004equilibrium,ruiz2021role}. In these models, naive KF-CK clusters gel (or percolate) above the critical temperature \cite{fajen2020percolation,charbonneau2021solution}, thus leaving the critical slowdown unaffected. 

Although various alternative approaches have been proposed, they are all somewhat unsatisfactory. Replica methods, which were first proposed by Swendsen and Wang \cite{swendsen1986replica} and developed into an algorithmic class that includes parallel tempering \cite{hukushima1996exchange,marinari1992simulated} and Houdayer's cluster algorithm \cite{houdayer2001cluster}, generically reduce the time needed to surmount reorganization barriers, but are computationally expensive because the number of replicas needed increases with system size. Single-replica approaches often lack generality in terms of model formulation, spatial dimension or lattice structure. We distinguish here three such approaches. 1) The Kandel-Ben-Av-Domany (KBD) cluster formalism and its variants \cite{kandel1990cluster,kandel1992cluster,cataudella1994critical,cataudella1996percolation}, which break up large frozen clusters using information from elementary plaquettes instead of spin pairs, are typically useful for certain square lattice models with competing nearest-neighbor interactions, but are inefficient for spin glass formers \cite{coddington1994generalized}. 
2) Dual worm (or loop) algorithms \cite{prokof2001worm,hitchcock2004dual}, which sample a dual lattice formulated in terms of bond variables, have been successfully generalized to fully frustrated models \cite{wang2012generalized} and spin glasses \cite{wang2005worm}, but are limited to 2D models with nearest-neighbor interactions. 3) Flat histogram methods, especially Wang-Landau sampling \cite{wang2001efficient}, which performs a random walk in energy space to overcome energy barriers and directly estimates density of states to obtain system properties, reduce but do not eliminate the critical slowdown of pure Ising models, and display correlation times that increases exponentially with size for frustrated systems \cite{dayal2004performance}. A more direct and generic cluster-like scheme would thus be desirable for these systems.

Pleimling and Henkel (PH) have proposed a cluster scheme for SALR models~\cite{pleimling2001anisotropic},
such as the archetypal axial next-nearest-neighbor Ising (ANNNI) model for spin variables $s_i=\pm 1$,
\begin{equation}
    \mathcal{H}_{\mathrm{ANNNI}} = -J\sum\limits_{\langle i,j\rangle}s_i s_j + \kappa J \sum\limits_{\lbrack i,j\rbrack_{\mathrm{axial}}}s_i s_j,
\end{equation}
where $J>0$ sets the unit of energy with SALR ($\kappa > 0$) and purely attractive ($\kappa < 0$) interactions.
Although the PH approach was reported to enhance sampling\cite{pleimling2001anisotropic,zhang2010monte}, its theoretical foundations and computational performance have not been previously assessed. It is therefore unclear to what extent it actually reduces the critical slowdown. In this Communication, we consider the PH scheme as well as modified cluster algorithms that better capture structural correlations in SALR models, so as to approach the algorithmic design target of Coniglio \cite{cataudella1994critical,cataudella1996percolation},
\begin{equation}
\label{eq:equivalence}
    \langle s_i s_j \rangle = \langle \gamma_{ij}\rangle,
\end{equation}
where  $\langle\gamma_{ij}\rangle$  denotes the probability that sites $i$ and $j$ are connected through a finite path and thus belong to the same cluster. Macroscopically, this condition results in the cluster percolation temperature coinciding with the model critical point. Although we do not here reach this objective, we nevertheless achieve efficiency gains of factors of up to forty in certain regimes of the ANNNI model.
The rest of this article first analytically considers the geometric and critical properties of the ANNNI model on 1D chains and on Bethe lattices, and then uses the resulting insights to 
formulate algorithms that are tested on a 2D square lattice version of the model. 

\begin{figure*}[t]
\centering
\includegraphics[scale=0.57,trim={0.4cm 0 0.1cm 0.2cm},clip]{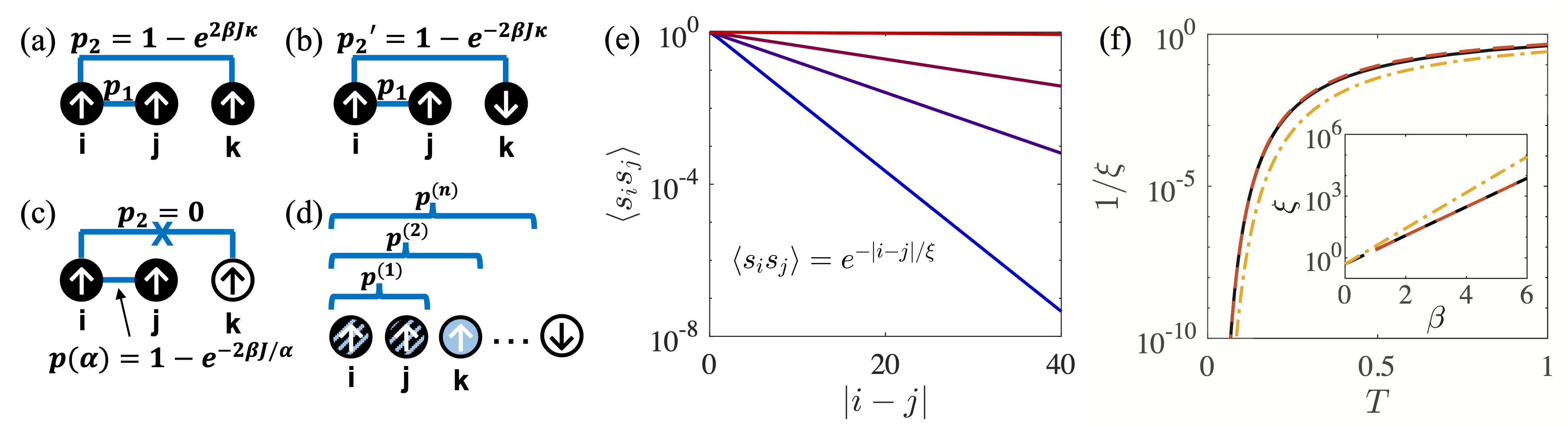}
\caption{Schematics for constructing (a) JCS, (b) PH, (c) modified Wolff and (d) TM clusters. Arrows denote spin orientations, and circles indicate whether a site belongs (black) or not (white) to the cluster of site $i$. In (a), the parallel nearest and next-nearest neighbors connect with probability $p_{1}$ and $p_{2}$, respectively. In (b), the parallel nearest neighbors are connected with probability $p_{1}$, and the antiparallel next-nearest neighbor with probability $p_{2}'$. In (c), only nearest neighbor connections are considered, with probability $p(\alpha)$ given by Eq.~\eqref{eq:modWolff_p}, where $\alpha$ is a tuning parameter. In (d), a site can be connected to $n$ contiguous parallel neighbor sites depending on the probability $p^{(n)}$ as given in Eq.~\eqref{eq:TM_pn}. (e) Decay of the spin-spin correlation $\langle s_i s_j \rangle$ with distance $\vert i-j\vert$ in the 1D chain for $\kappa=0.1$ and $\beta = 1, 1.5, 2, 4$ from bottom to top. (f) The correlation length (black solid line) $\xi$ diverges as $T\rightarrow 0$, which is obtained from the exponential fitting of (e). The inset shows the scaling $\xi = e^{(1-2\kappa)2\beta J}/2$ with the $\beta\rightarrow0$ extrapolation obtained by fitting. Cluster sizes also diverge at $T\rightarrow 0$. JCS clusters (orange dash line) scale similarly as the correlation length, whereas PH clusters (yellow dot-dash line) do not. (Without loss of generality, only results for $\kappa=0.1$ are shown.)}
\label{fig:annni1d}
\end{figure*}

\paragraph*{1D Chain --} One-dimensional models can be solved using transfer matrices, and are thus helpful in obtaining insights into attaining Eq.~\eqref{eq:equivalence}. First, spin-spin correlations a distance $r$ apart can be expressed as
\begin{equation}
    \langle s_i s_{i+r} \rangle = \frac{1}{Z} \sum\limits_\nu s_i s_{i+r} e^{-\beta H_\nu}
    = \frac{1}{Z} \mathrm{Tr}[\sigma_z T^r \sigma_z T^{N-r}]
\end{equation}
where the $4\times 4$ transfer matrix $\mathbf{T}=\exp{(s_i (s_{i+1} -\kappa s_{i+2})\beta J )}$ is sparse with eight nonzero entries, $\sigma_z = \mathrm{Diag}(1,1,-1,-1)$, and the partition function, $Z = \mathrm{Tr}[T^N]$.
Second, site-bond correlated percolation can be obtained by generalizing the approach of Derrida\cite{derrida1980TM}. Let the vector $P(r+1)$ denote the probability distribution that site $r+1$ is connected to site $1$ through a finite path, depending on the connection probability distribution of backward sites $P(r)$, and iteratively write
\begin{equation}
    P(r+1) = M P(r) = M^r P(1),
\end{equation}
where $M$ is a matrix that contains Boltzmann weight and bonding probabilities and $P(1)$ is actually the configuration distribution for the rest $N-r$ sites under periodic boundary conditions. The connection probability of two sites a distance $r$ apart is generally expressed as $\langle \gamma_{i,i+r} \rangle = \mathrm{Tr}[M^r T^{N-r}]/Z$, but requires correction terms when next-nearest-neighbor bonds are considered~\cite{SI}.
Because both the orientation and the \textit{cluster connection} of a spin need to be taken into account, $M$ is generically a $16\times 16$ sparse matrix, but its size can be reduced for specific cluster definitions. In particular, for standard Jan-Coniglio-Stauffer (JCS) clusters~\cite{jan1982study}, in which only parallel sites are potentially connected,  using a bonding probability $p_1 = 1 - e^{-2\beta J}$ for nearest neighbors and $p_2 = 1 - e^{2\beta J\kappa}$ for next-nearest neighbors (see Fig.~\ref{fig:annni1d}(c)) gives
\begin{widetext}
\begin{table}[h]
    \raggedright
    \begin{tabular}{ccc}
       &  & 
      \begin{minipage}[b]{1.8\columnwidth}
      \centering
      \raisebox{-.5\height}{
      \begin{minipage}[b]{0.27\columnwidth}
      {\includegraphics[scale=0.35]{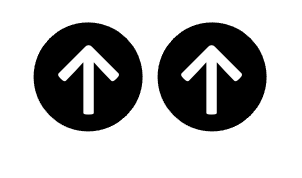}}
      \end{minipage}
      \begin{minipage}[b]{0.24\columnwidth}
      {\includegraphics[scale=0.35]{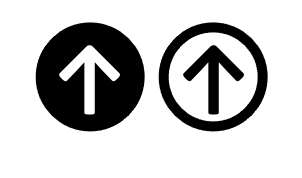}}
      \end{minipage}
      \begin{minipage}[b]{0.16\columnwidth}
      {\includegraphics[scale=0.35]{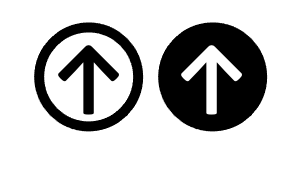}}
      \end{minipage}
      \begin{minipage}[b]{0.13\columnwidth}
      {\includegraphics[scale=0.35]{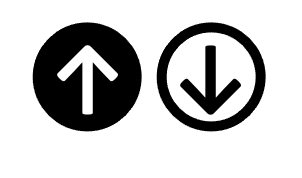}}
      \end{minipage}
      \begin{minipage}[b]{0.2\columnwidth}
      {\includegraphics[scale=0.35]{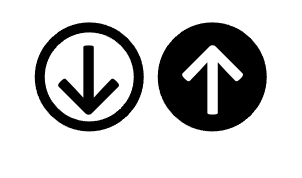}\qquad\qquad\qquad}
      \end{minipage}
      }
      \end{minipage}
      \\
     $M=$ & 
     \begin{minipage}[b]{0.12\columnwidth}
      \centering
      \raisebox{-.5\height}{
      \includegraphics[scale=0.35]{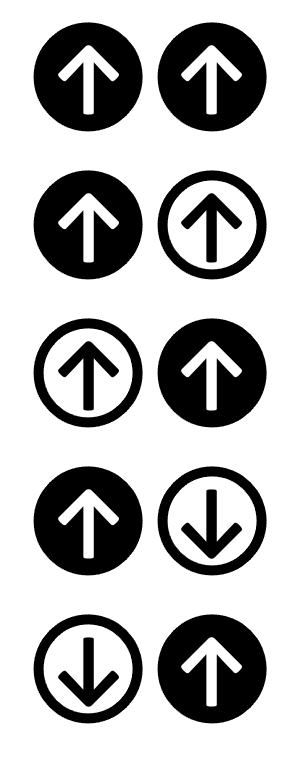}
      }
      \end{minipage}
     & 
     \begin{minipage}[b]{1.8\columnwidth}
      \centering
      \begin{equation}
      \left( \begin{array}{ccccc}
    (p_1+p_2-p_1p_2)e^{\beta J(1-\kappa)} & (1-p_1)(1-p_2)e^{\beta J(1-\kappa)} & 0 & e^{\beta J(1+\kappa)} & 0 \\
    p_1p_2e^{\beta J(1-\kappa)} & 0 & (1-p_1)p_2e^{\beta J(1-\kappa)} & 0 & 0 \\
    p_1e^{\beta J(1-\kappa)} & (1-p_1)e^{\beta J(1-\kappa)} & 0 & e^{\beta J(1+\kappa)} & 0 \\
    0 & 0 & 0 & 0 & p_2e^{-\beta J(1+\kappa)} \\
    p_1e^{-\beta J(1-\kappa)} & (1-p_1)e^{-\beta J(1-\kappa)} & 0 & e^{-\beta J(1+\kappa)} & 0
     \end{array} \right),
     \label{eq:NNNA_M}
    \end{equation}
      \end{minipage}
    \end{tabular}
    \label{tab:my_label}
\end{table}
\end{widetext}
where each line encodes a specific configuration for sites $i$ and $i+1$, and each column for $i+1$ and $i+2$. Arrows denote spin orientations, and colors encode connection to site $1$ (black) or not (white).

In the purely attractive ($\kappa<0$) case, Eq.~\eqref{eq:equivalence} holds for JCS clusters (and for all other purely attractive models \cite{jan1982study}), as is here verified. Surprisingly, this equivalence persists even in the SALR regime, albeit for a \textit{negative bonding probability}, i.e., $p_2 < 0$. It is noticeable that the matrix elements $m_{21}$, $m_{23}$ and $m_{45}$ are thus negative, indicating that the existence of next-nearest neighbors with repulsion in the cluster reduces $\langle\gamma_{ij}\rangle$. By contrast, PH clusters, which connect parallel nearest neighbors with probability $p_1$ and antiparallel next-nearest neighbors with probability $p_2' = 1 - e^{2\beta\kappa J} > 0$ (see Fig.~\ref{fig:annni1d}(d)), 
have $\langle\gamma_{ij}\rangle>\langle s_i s_j \rangle$ for all $\kappa>0$. For such a 1D model, we nevertheless have that $T_c=T_p=0$ for both schemes. Cluster sizes of both types therefore diverge along with the correlation length (see Fig.~\ref{fig:annni1d}(e)(f)), which scales as 
\begin{equation}
\label{eq:annni1d_scaling}
    \xi = e^{(1-2\kappa)2\beta J}/2.
\end{equation} 
Although JCS cluster sizes scale accordingly, PH clusters do not. This discrepancy hints at a possible inadequacy of the latter scheme, but further insights are needed to understand why. 

\begin{figure}[t]
\centering
\subfigure{
\begin{minipage}{\columnwidth}
\centering
\includegraphics[scale=0.42,trim={0.2cm 0 0.2cm 0.2cm},clip]{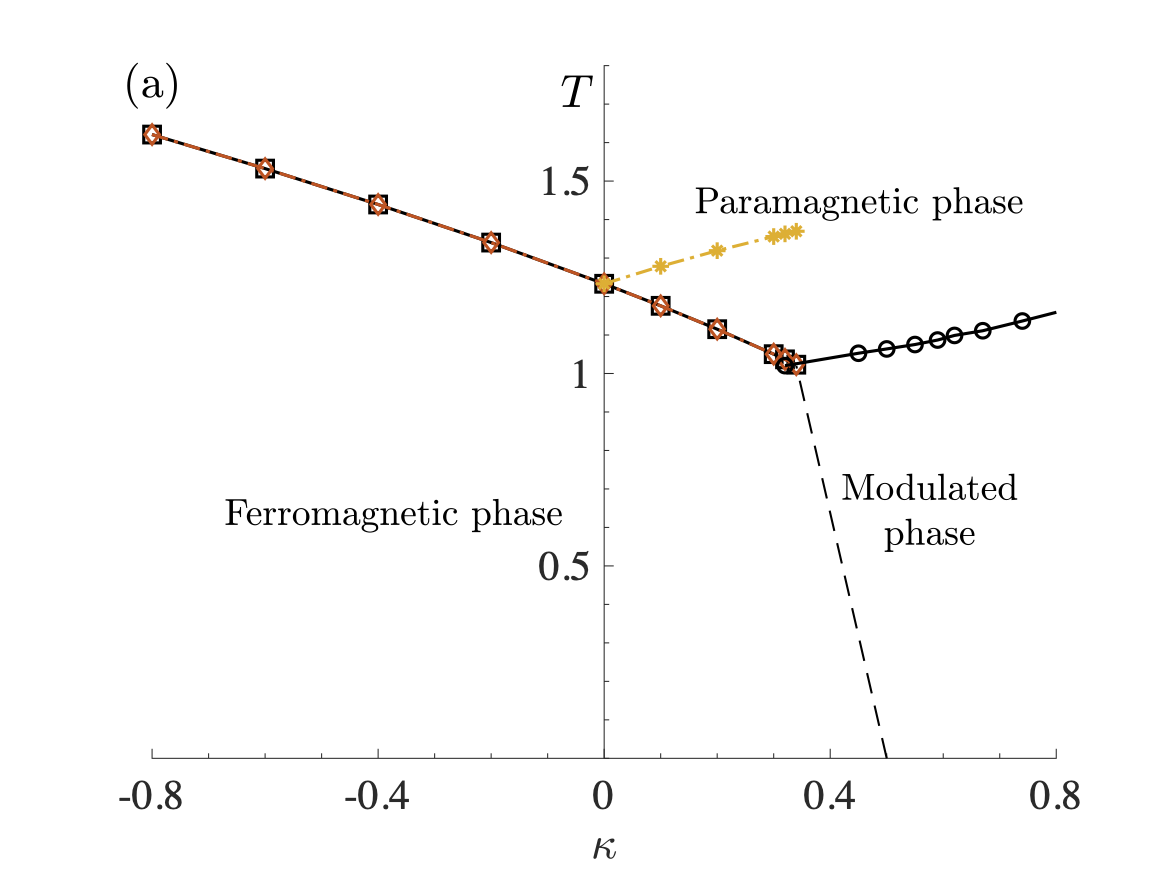}
\includegraphics[scale=0.42,trim={0.2cm 0 0.2cm 0.2cm},clip]{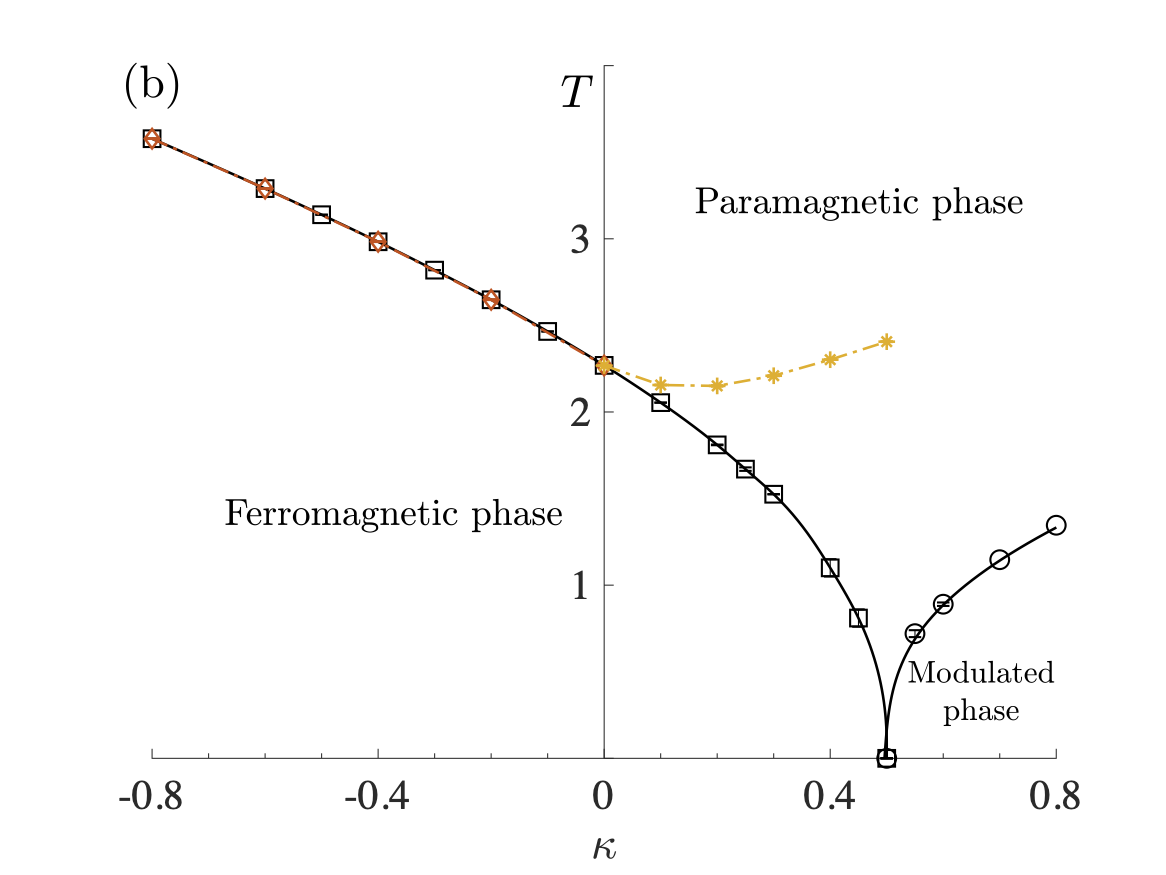}
\end{minipage}
}
\caption{Phase diagram of the ANNNI model on (a)  a Bethe lattice with connectivity $c+1=6$, and (b) a 2D square lattice (adapted and extended~\cite{SI} from Ref.~\onlinecite[Fig.~3]{hu2021resolving}). The percolation temperature for JCS clusters (orange diamonds) agrees with $T_c$ (black squares), while for PH clusters (yellow stars) $T_p>T_c$. Note that the paramagnetic-ferromagnetic transition is only observed for $\kappa<\kappa_L$, with Lifshitz point in (a) $\kappa_L = 0.34$ and in (b) $\kappa_L = 0.5$. In (a), the boundary between the ferromagnetic and modulated phases is estimated by linearly interpolating between the $\kappa_L$ and the $T=0$ ground state $\kappa_0=0.5$ (dashed line).}
\label{fig:annni_phase}
\end{figure}

\paragraph*{Bethe Lattice --} We therefore consider the ANNNI model on a Bethe lattice, which offers (non-ambiguous) non-zero transition temperatures and is also analytically solvable, albeit through the slightly more involved cavity field method \cite{semerjian2009exact,charbonneau2021solution,zmc2022}. The resulting phase diagram and percolation temperatures of the JCS clusters as well the PH clusters are shown in Fig.~\ref{fig:annni_phase}(a). In the paramagnetic-ferromagnetic phase regime (for $\kappa < \kappa_L$, where $\kappa_L$ is the Lifshitz point beyond which the ferromagnetic ground state is replaced by a modulated one), the percolation temperature of JCS clusters perfectly coincides with the critical one, as long as \emph{a negative bonding probability} is here again used for $\kappa>0$. The PH clusters, by contrast, percolate at temperatures that grow more distant from the critical point as $\kappa$ increases. The percolating, gel-like clusters are then expected to struggle to sample configurations in the critical regime. Interestingly, this analysis further reveals that  spin-spin correlations in this regime do not include spins of opposite sign  despite the presence of an antiferromagnetic coupling.

Given the clear importance of negative probabilities (or rather a diminution of binding probabilities) in JCS clusters, we propose two algorithms -- the modified Wolff algorithm (Fig.~\ref{fig:annni1d}(c)) and the Transfer Matrix (TM) cluster algorithm (Fig.~\ref{fig:annni1d}(d)) -- to at least partially account for this effect. In the modified Wolff algorithm, the connection between the parallel nearest neighbors has a bonding probability 
\begin{equation}
\label{eq:modWolff_p}
    p(\alpha) = 1 - e^{-2\beta J/\alpha}
\end{equation}
tuned to attain $T_c = T_p$, for which the algorithm efficiency is optimal. The contribution of next-nearest neighbors should increase $p(\alpha)$ for purely attractive interactions and decrease $p(\alpha)$ for the SALR regime, thus resulting in diverging cluster sizes at the critical point. In the TM cluster algorithm, the initial site $i$ is connected to the next $n$ contiguous parallel neighbors in a randomly selected direction with probability $p^{(n)}$ given by the top-left $2\times 2$ block of Eq.~\eqref{eq:NNNA_M}\cite{SI},
\begin{equation}
\begin{split}
    (r_{n+1}, \ s_{n+1}) = a M_0^{n} 
    = (p_1, 1-p_1)\left( \begin{array}{cc}
    m_{11} & m_{12} \\
    m_{21} & m_{22}
     \end{array} \right)^{n},
\end{split}
\label{eq:TM_pn}
\end{equation}
\begin{equation}
    p^{(n)} = \left\{ \begin{array}{ll}
              r_m & \mbox{if $n=m$};\\
              s_{n+1} & \mbox{if $0 < n < m$};\\
1-\sum\limits_{i=1}^m s_i - r_m & \mbox{if $n = 0$}.\end{array} \right.
\end{equation}
where $s_{n+1}$ denotes the probability that growth of the cluster stops at the $n$-th neighbor spin, and $r_{n+1}$ that clustering continues, given the maximum number of contiguous parallel neighbors equals to $m$. Because $M_0$ is but a sub-matrix of $M$, the summation of all considered clustering probabilities does not add up to unity, so we assign the residual probability to the case $n=0$, which  should not affect the algorithmic performance. Note that this partial correction of the negative contribution of next-nearest neighbors could be extended to a larger subset of $M$, but the algorithmic complexity would then grow significantly and is thus left for future consideration. 

\paragraph*{2D Square Lattice --} In order to assess the performance of these algorithms, we consider a 2D square lattice version of the ANNNI model. By extending the TM approach of Ref.~\onlinecite{hu2021resolving}, a high-precision phase diagram of the model is obtained (Fig.~\ref{fig:annni_phase}(b)), and numerical simulations along with finite-size scaling \cite{cataudella1992percolation} are used determine the percolation temperatures of JCS clusters for $\kappa < 0$ and PH clusters for $\kappa > 0$. The resulting phase and percolation behaviors are qualitatively similar to those of the Bethe lattice version.

MC simulations are then performed on the same lattices with $L^2$ spins. In order to extract a correlation (mixing) time $\tau$ for the various algorithms, we consider the decay the autocorrelation function of the absolute magnetization $\lvert m\rvert$~\cite{baillie1991comparison,cataudella1996percolation},
\begin{equation}
    C(t) = \frac{\langle\lvert m(0)m(t)\rvert\rangle-\langle\lvert m(0)\rvert\rangle^2}{\langle\lvert m(0)^2\rvert\rangle-\langle \lvert m(0)\rvert\rangle^2}
\end{equation}
which is fitted using a (stretched) exponential function \cite{ediger2000spatially,bouchaud2008anomalous}
\begin{equation}
    C(t) \propto \exp{[-(t/\tau)^\gamma]},
\end{equation}
where $0 < \gamma < 1$. When appropriate, the critical dynamical scaling form, $\tau = \tau_0 L^z$ with dynamical exponent $z$ is used to fit the results \cite{newman1999monte}. For the sake of comparison, time units here generally denote $N=L^2$ spins having been considered for a flip. In particular, for standard single-spin-flip (SSF) algorithm, $t = t_{\mathrm{step}}/L^2$; for any cluster algorithms, $t = t_{\mathrm{step}}\langle n \rangle / L^2$, where $t_{\mathrm{step}}$ is the number of MC steps, and $\langle n \rangle$ is the mean cluster size generated by the algorithm \cite{tamayo1990single}. The resulting $\tau$ are then proportional to their CPU time with only negligible finite-size differences. Initial equilibration times are at least $10\tau$, and sampling times at least $100\tau$. 

\begin{figure*}[t]
\centering
\subfigure{
\begin{minipage}{\textwidth}
\centering
\includegraphics[scale=0.58,trim={0.1cm 0.08cm 0.4cm 0.4cm},clip]{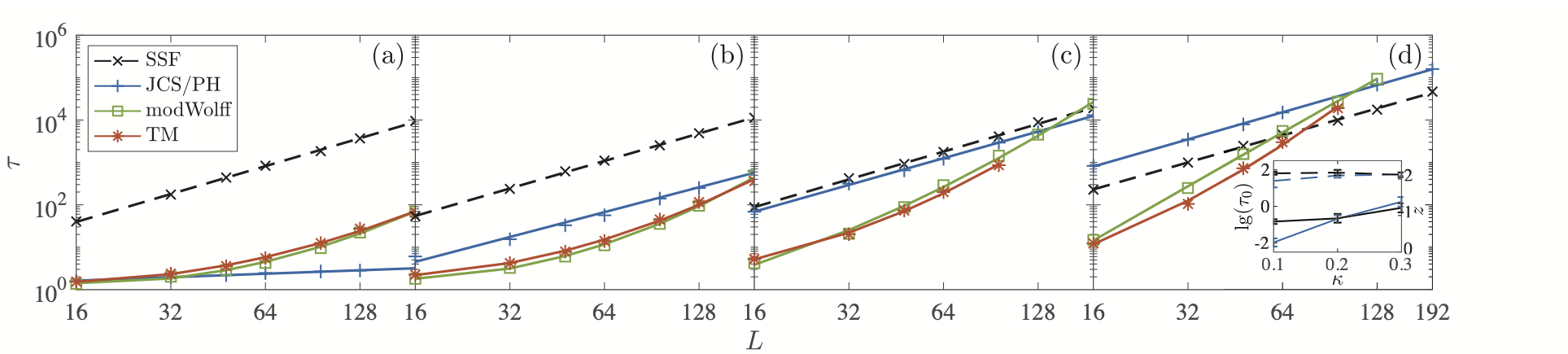}
\end{minipage}
}
\caption{Mixing time $\tau$ for various simulation algorithms applied to the ANNNI model on a 2D square lattice at $T_c$ for (a) $\kappa=-0.1$, (b) $\kappa=0.1$, (c) $\kappa=0.2$ and (d) $\kappa=0.3$. The standard SSF algorithm is roughly invariant to changing $\kappa$ (inset). The JCS cluster algorithm largely eliminates the critical slowdown for purely attractive systems, but its PH rejection-free counterpart for $\kappa>0$ scales like the SSF algorithm albeit with a smaller prefactor at low $\kappa$ (see the inset). The proposed cluster schemes are most efficient at small $L$.}
\label{fig:annni2d_corrtime}
\end{figure*}

The growth of $\tau$ with $L$ for various cluster schemes is shown in Fig.~\ref{fig:annni2d_corrtime}, along with SSF results, which are roughly invariant of $\kappa$ and are therefore used as reference. As expected, for purely attractive systems the JCS cluster algorithm is most efficient, with a dynamical exponent as small as that of the standard Wolff (CK-FK cluster-based) algorithm for the simple Ising model \cite{wolff1989collective}. Although the two proposed algorithms are less efficient, they are nevertheless significantly more so than SSF. By construction, the modified Wolff algorithm generates clusters that percolate at $T_c$, but unlike the JCS clusters, this scheme is not rejection-free. The rejection rate therefore necessarily increases as the cluster boundary (or energy cost) increases with $L$. The TM cluster algorithm only includes part of the next-nearest-neighbor contribution to correlations and therefore the constructed clusters are not ideal at decorrelating configurations, but its performance is nevertheless very similar to that of the modified Wolff algorithm. The TM algorithm additionally benefits from its parameter-free nature.

In the SALR regime, the PH cluster algorithm exhibits the same dynamical scaling as SSF, but with a markedly reduced prefactor $\tau_0$ at weak $\kappa$. This efficiency gain probably explains why earlier reports found the algorithm helpful 
\cite{pleimling2001anisotropic,zhang2010monte}. As $\kappa$ increases, however, that prefactor grows rapidly (see Fig.~\ref{fig:annni2d_corrtime} inset). The considerable computational cost of building clusters then makes it lose out even to the SSF algorithm. By contrast, the two proposed algorithms exhibit prefactors that scale more favorably with $\kappa$, albeit not the critical exponents. They are therefore most efficient for small system sizes. Because the mixing times grow either polynomially or exponentially---given that the rejection rate increases with $L$---the approaches become less advantageous as $L$ grows.

\paragraph*{Conclusion --}
Based on a first-principle study of correlations and clustering, we have proposed two algorithms that account for the negative bonding probability of the next-nearest neighbors in models with SALR interactions. These approaches are reminiscent of the KBD algorithm, but are adapted to systems for which the standard plaquettes-sharing bonds would normally fail. Unlike the PH algorithm, the schemes also approximate the Coniglio criterion (Eq.~\eqref{eq:equivalence}). They therefore present significant computational gains, albeit only over a limited range of system sizes, given their non-rejection-free nature. Slightly away from the critical point, they should therefore be of great assistance, but a rejection-free version of these scheme would be needed to improve the configurational sampling of SALR and  other frustrated models more generally. Adapting these algorithms to conditions in which the ground state is not ferromagnetic is another important future direction.

\paragraph*{Acknowledgments -- }
We thank Ye Liang for sharing unpublished results and Yi Hu for sharing his TM code as well as for various discussions. We also than Antonio Coniglio for motivating exchanges. We acknowledge support from the Simons Foundation (Grant No. 454937) and from the National Science Foundation, Grant No. DMR-1749374. Computations were carried out on the Duke Compute Cluster. Data associated with this work are available from the Duke Digital Repository\cite{SALRdata}.

\bibliography{reference}

\begin{thebibliography}{51}%
\makeatletter
\providecommand \@ifxundefined [1]{%
 \@ifx{#1\undefined}
}%
\providecommand \@ifnum [1]{%
 \ifnum #1\expandafter \@firstoftwo
 \else \expandafter \@secondoftwo
 \fi
}%
\providecommand \@ifx [1]{%
 \ifx #1\expandafter \@firstoftwo
 \else \expandafter \@secondoftwo
 \fi
}%
\providecommand \natexlab [1]{#1}%
\providecommand \enquote  [1]{``#1''}%
\providecommand \bibnamefont  [1]{#1}%
\providecommand \bibfnamefont [1]{#1}%
\providecommand \citenamefont [1]{#1}%
\providecommand \href@noop [0]{\@secondoftwo}%
\providecommand \href [0]{\begingroup \@sanitize@url \@href}%
\providecommand \@href[1]{\@@startlink{#1}\@@href}%
\providecommand \@@href[1]{\endgroup#1\@@endlink}%
\providecommand \@sanitize@url [0]{\catcode `\\12\catcode `\$12\catcode
  `\&12\catcode `\#12\catcode `\^12\catcode `\_12\catcode `\%12\relax}%
\providecommand \@@startlink[1]{}%
\providecommand \@@endlink[0]{}%
\providecommand \url  [0]{\begingroup\@sanitize@url \@url }%
\providecommand \@url [1]{\endgroup\@href {#1}{\urlprefix }}%
\providecommand \urlprefix  [0]{URL }%
\providecommand \Eprint [0]{\href }%
\providecommand \doibase [0]{http://dx.doi.org/}%
\providecommand \selectlanguage [0]{\@gobble}%
\providecommand \bibinfo  [0]{\@secondoftwo}%
\providecommand \bibfield  [0]{\@secondoftwo}%
\providecommand \translation [1]{[#1]}%
\providecommand \BibitemOpen [0]{}%
\providecommand \bibitemStop [0]{}%
\providecommand \bibitemNoStop [0]{.\EOS\space}%
\providecommand \EOS [0]{\spacefactor3000\relax}%
\providecommand \BibitemShut  [1]{\csname bibitem#1\endcsname}%
\let\auto@bib@innerbib\@empty
\bibitem [{\citenamefont {Newman}\ and\ \citenamefont
  {Barkema}(1999)}]{newman1999monte}%
  \BibitemOpen
  \bibfield  {author} {\bibinfo {author} {\bibfnamefont {M.~E.~J.}\
  \bibnamefont {Newman}}\ and\ \bibinfo {author} {\bibfnamefont {G.~T.}\
  \bibnamefont {Barkema}},\ }\href@noop {} {\emph {\bibinfo {title} {Monte
  Carlo methods in statistical physics}}}\ (\bibinfo  {publisher} {Clarendon
  Press},\ \bibinfo {year} {1999})\BibitemShut {NoStop}%
\bibitem [{\citenamefont {Frenkel}\ and\ \citenamefont
  {Smit}(2001)}]{frenkel2001understanding}%
  \BibitemOpen
  \bibfield  {author} {\bibinfo {author} {\bibfnamefont {D.}~\bibnamefont
  {Frenkel}}\ and\ \bibinfo {author} {\bibfnamefont {B.}~\bibnamefont {Smit}},\
  }\href@noop {} {\emph {\bibinfo {title} {Understanding molecular simulation:
  from algorithms to applications}}},\ Vol.~\bibinfo {volume} {1}\ (\bibinfo
  {publisher} {Elsevier},\ \bibinfo {year} {2001})\BibitemShut {NoStop}%
\bibitem [{\citenamefont {Landau}\ and\ \citenamefont
  {Binder}(2021)}]{landau2021guide}%
  \BibitemOpen
  \bibfield  {author} {\bibinfo {author} {\bibfnamefont {D.}~\bibnamefont
  {Landau}}\ and\ \bibinfo {author} {\bibfnamefont {K.}~\bibnamefont
  {Binder}},\ }\href@noop {} {\emph {\bibinfo {title} {A guide to Monte Carlo
  simulations in statistical physics}}}\ (\bibinfo  {publisher} {Cambridge
  university press},\ \bibinfo {year} {2021})\BibitemShut {NoStop}%
\bibitem [{\citenamefont {Niedermayer}(1988)}]{niedermayer1988general}%
  \BibitemOpen
  \bibfield  {author} {\bibinfo {author} {\bibfnamefont {F.}~\bibnamefont
  {Niedermayer}},\ }\href@noop {} {\bibfield  {journal} {\bibinfo  {journal}
  {Phys. Rev. Lett.}\ }\textbf {\bibinfo {volume} {61}},\ \bibinfo {pages}
  {2026} (\bibinfo {year} {1988})}\BibitemShut {NoStop}%
\bibitem [{\citenamefont {Edwards}\ and\ \citenamefont
  {Sokal}(1988)}]{edwards1988generalization}%
  \BibitemOpen
  \bibfield  {author} {\bibinfo {author} {\bibfnamefont {R.~G.}\ \bibnamefont
  {Edwards}}\ and\ \bibinfo {author} {\bibfnamefont {A.~D.}\ \bibnamefont
  {Sokal}},\ }\href@noop {} {\bibfield  {journal} {\bibinfo  {journal} {Phys.
  Rev. D}\ }\textbf {\bibinfo {volume} {38}},\ \bibinfo {pages} {2009}
  (\bibinfo {year} {1988})}\BibitemShut {NoStop}%
\bibitem [{\citenamefont {Machta}\ \emph {et~al.}(1995)\citenamefont {Machta},
  \citenamefont {Choi}, \citenamefont {Lucke}, \citenamefont {Schweizer},\ and\
  \citenamefont {Chayes}}]{machta1995invaded}%
  \BibitemOpen
  \bibfield  {author} {\bibinfo {author} {\bibfnamefont {J.}~\bibnamefont
  {Machta}}, \bibinfo {author} {\bibfnamefont {Y.~S.}\ \bibnamefont {Choi}},
  \bibinfo {author} {\bibfnamefont {A.}~\bibnamefont {Lucke}}, \bibinfo
  {author} {\bibfnamefont {T.}~\bibnamefont {Schweizer}}, \ and\ \bibinfo
  {author} {\bibfnamefont {L.~V.}\ \bibnamefont {Chayes}},\ }\href@noop {}
  {\bibfield  {journal} {\bibinfo  {journal} {Phys. Rev. Lett.}\ }\textbf
  {\bibinfo {volume} {75}},\ \bibinfo {pages} {2792} (\bibinfo {year}
  {1995})}\BibitemShut {NoStop}%
\bibitem [{\citenamefont {Whitelam}\ and\ \citenamefont
  {Geissler}(2007)}]{whitelam2007avoiding}%
  \BibitemOpen
  \bibfield  {author} {\bibinfo {author} {\bibfnamefont {S.}~\bibnamefont
  {Whitelam}}\ and\ \bibinfo {author} {\bibfnamefont {P.~L.}\ \bibnamefont
  {Geissler}},\ }\href@noop {} {\bibfield  {journal} {\bibinfo  {journal} {J.
  Chem. Phys.}\ }\textbf {\bibinfo {volume} {127}},\ \bibinfo {pages} {154101}
  (\bibinfo {year} {2007})}\BibitemShut {NoStop}%
\bibitem [{\citenamefont {R\u{u}{\v{z}}i{\v{c}}ka}\ and\ \citenamefont
  {Allen}(2014)}]{ruuvzivcka2014collective}%
  \BibitemOpen
  \bibfield  {author} {\bibinfo {author} {\bibfnamefont {{\v{S}}.}~\bibnamefont
  {R\u{u}{\v{z}}i{\v{c}}ka}}\ and\ \bibinfo {author} {\bibfnamefont {M.~P.}\
  \bibnamefont {Allen}},\ }\href@noop {} {\bibfield  {journal} {\bibinfo
  {journal} {Phys. Rev. E}\ }\textbf {\bibinfo {volume} {90}},\ \bibinfo
  {pages} {033302} (\bibinfo {year} {2014})}\BibitemShut {NoStop}%
\bibitem [{\citenamefont {Chen}\ and\ \citenamefont
  {Siepmann}(2000)}]{chen2000novel}%
  \BibitemOpen
  \bibfield  {author} {\bibinfo {author} {\bibfnamefont {B.}~\bibnamefont
  {Chen}}\ and\ \bibinfo {author} {\bibfnamefont {J.~I.}\ \bibnamefont
  {Siepmann}},\ }\href@noop {} {\bibfield  {journal} {\bibinfo  {journal} {J.
  Phys. Chem. B}\ }\textbf {\bibinfo {volume} {104}},\ \bibinfo {pages} {8725}
  (\bibinfo {year} {2000})}\BibitemShut {NoStop}%
\bibitem [{\citenamefont {Michel}, \citenamefont {Kapfer},\ and\ \citenamefont
  {Krauth}(2014)}]{michel2014generalized}%
  \BibitemOpen
  \bibfield  {author} {\bibinfo {author} {\bibfnamefont {M.}~\bibnamefont
  {Michel}}, \bibinfo {author} {\bibfnamefont {S.~C.}\ \bibnamefont {Kapfer}},
  \ and\ \bibinfo {author} {\bibfnamefont {W.}~\bibnamefont {Krauth}},\
  }\href@noop {} {\bibfield  {journal} {\bibinfo  {journal} {J. Chem. Phys.}\
  }\textbf {\bibinfo {volume} {140}},\ \bibinfo {pages} {054116} (\bibinfo
  {year} {2014})}\BibitemShut {NoStop}%
\bibitem [{\citenamefont {Krauth}(2021)}]{krauth2021event}%
  \BibitemOpen
  \bibfield  {author} {\bibinfo {author} {\bibfnamefont {W.}~\bibnamefont
  {Krauth}},\ }\href@noop {} {\bibfield  {journal} {\bibinfo  {journal} {Front.
  Phys.}\ }\textbf {\bibinfo {volume} {9}},\ \bibinfo {pages} {663457}
  (\bibinfo {year} {2021})}\BibitemShut {NoStop}%
\bibitem [{\citenamefont {Kasteleyn}\ and\ \citenamefont
  {Fortuin}(1969)}]{kasteleyn1969phase}%
  \BibitemOpen
  \bibfield  {author} {\bibinfo {author} {\bibfnamefont {P.~W.}\ \bibnamefont
  {Kasteleyn}}\ and\ \bibinfo {author} {\bibfnamefont {C.~M.}\ \bibnamefont
  {Fortuin}},\ }\href@noop {} {\bibfield  {journal} {\bibinfo  {journal} {J.
  Phys. Soc. Japan}\ }\textbf {\bibinfo {volume} {26}},\ \bibinfo {pages} {11}
  (\bibinfo {year} {1969})}\BibitemShut {NoStop}%
\bibitem [{\citenamefont {Fortuin}\ and\ \citenamefont
  {Kasteleyn}(1972)}]{fortuin1972random}%
  \BibitemOpen
  \bibfield  {author} {\bibinfo {author} {\bibfnamefont {C.~M.}\ \bibnamefont
  {Fortuin}}\ and\ \bibinfo {author} {\bibfnamefont {P.~W.}\ \bibnamefont
  {Kasteleyn}},\ }\href@noop {} {\bibfield  {journal} {\bibinfo  {journal}
  {Physica}\ }\textbf {\bibinfo {volume} {57}},\ \bibinfo {pages} {536}
  (\bibinfo {year} {1972})}\BibitemShut {NoStop}%
\bibitem [{\citenamefont {Coniglio}\ and\ \citenamefont
  {Klein}(1980)}]{coniglio1980clusters}%
  \BibitemOpen
  \bibfield  {author} {\bibinfo {author} {\bibfnamefont {A.}~\bibnamefont
  {Coniglio}}\ and\ \bibinfo {author} {\bibfnamefont {W.}~\bibnamefont
  {Klein}},\ }\href@noop {} {\bibfield  {journal} {\bibinfo  {journal} {J.
  Phys. A}\ }\textbf {\bibinfo {volume} {13}},\ \bibinfo {pages} {2775}
  (\bibinfo {year} {1980})}\BibitemShut {NoStop}%
\bibitem [{\citenamefont {Swendsen}\ and\ \citenamefont
  {Wang}(1987)}]{swendsen1987nonuniversal}%
  \BibitemOpen
  \bibfield  {author} {\bibinfo {author} {\bibfnamefont {R.~H.}\ \bibnamefont
  {Swendsen}}\ and\ \bibinfo {author} {\bibfnamefont {J.-S.}\ \bibnamefont
  {Wang}},\ }\href@noop {} {\bibfield  {journal} {\bibinfo  {journal} {Phys.
  Rev. Lett.}\ }\textbf {\bibinfo {volume} {58}},\ \bibinfo {pages} {86}
  (\bibinfo {year} {1987})}\BibitemShut {NoStop}%
\bibitem [{\citenamefont {Wolff}(1989)}]{wolff1989collective}%
  \BibitemOpen
  \bibfield  {author} {\bibinfo {author} {\bibfnamefont {U.}~\bibnamefont
  {Wolff}},\ }\href@noop {} {\bibfield  {journal} {\bibinfo  {journal} {Phys.
  Rev. Lett.}\ }\textbf {\bibinfo {volume} {62}},\ \bibinfo {pages} {361}
  (\bibinfo {year} {1989})}\BibitemShut {NoStop}%
\bibitem [{\citenamefont {Binder}\ and\ \citenamefont
  {Young}(1986)}]{binder1986spin}%
  \BibitemOpen
  \bibfield  {author} {\bibinfo {author} {\bibfnamefont {K.}~\bibnamefont
  {Binder}}\ and\ \bibinfo {author} {\bibfnamefont {A.~P.}\ \bibnamefont
  {Young}},\ }\href@noop {} {\bibfield  {journal} {\bibinfo  {journal} {Rev.
  Mod. Phys.}\ }\textbf {\bibinfo {volume} {58}},\ \bibinfo {pages} {801}
  (\bibinfo {year} {1986})}\BibitemShut {NoStop}%
\bibitem [{\citenamefont {Langer}(2014)}]{langer2014theories}%
  \BibitemOpen
  \bibfield  {author} {\bibinfo {author} {\bibfnamefont {J.~S.}\ \bibnamefont
  {Langer}},\ }\href@noop {} {\bibfield  {journal} {\bibinfo  {journal} {Rep.
  Prog. Phys.}\ }\textbf {\bibinfo {volume} {77}},\ \bibinfo {pages} {042501}
  (\bibinfo {year} {2014})}\BibitemShut {NoStop}%
\bibitem [{\citenamefont {Stradner}\ \emph {et~al.}(2004)\citenamefont
  {Stradner}, \citenamefont {Sedgwick}, \citenamefont {Cardinaux},
  \citenamefont {Poon}, \citenamefont {Egelhaaf},\ and\ \citenamefont
  {Schurtenberger}}]{stradner2004equilibrium}%
  \BibitemOpen
  \bibfield  {author} {\bibinfo {author} {\bibfnamefont {A.}~\bibnamefont
  {Stradner}}, \bibinfo {author} {\bibfnamefont {H.}~\bibnamefont {Sedgwick}},
  \bibinfo {author} {\bibfnamefont {F.}~\bibnamefont {Cardinaux}}, \bibinfo
  {author} {\bibfnamefont {W.~C.~K.}\ \bibnamefont {Poon}}, \bibinfo {author}
  {\bibfnamefont {S.~U.}\ \bibnamefont {Egelhaaf}}, \ and\ \bibinfo {author}
  {\bibfnamefont {P.}~\bibnamefont {Schurtenberger}},\ }\href@noop {}
  {\bibfield  {journal} {\bibinfo  {journal} {Nature}\ }\textbf {\bibinfo
  {volume} {432}},\ \bibinfo {pages} {492} (\bibinfo {year}
  {2004})}\BibitemShut {NoStop}%
\bibitem [{\citenamefont {Ruiz-Franco}\ and\ \citenamefont
  {Zaccarelli}(2021)}]{ruiz2021role}%
  \BibitemOpen
  \bibfield  {author} {\bibinfo {author} {\bibfnamefont {J.}~\bibnamefont
  {Ruiz-Franco}}\ and\ \bibinfo {author} {\bibfnamefont {E.}~\bibnamefont
  {Zaccarelli}},\ }\href@noop {} {\bibfield  {journal} {\bibinfo  {journal}
  {Annu. Rev. Condens. Matter Phys}\ }\textbf {\bibinfo {volume} {12}},\
  \bibinfo {pages} {51} (\bibinfo {year} {2021})}\BibitemShut {NoStop}%
\bibitem [{\citenamefont {Fajen}, \citenamefont {Hartmann},\ and\ \citenamefont
  {Young}(2020)}]{fajen2020percolation}%
  \BibitemOpen
  \bibfield  {author} {\bibinfo {author} {\bibfnamefont {H.}~\bibnamefont
  {Fajen}}, \bibinfo {author} {\bibfnamefont {A.~K.}\ \bibnamefont {Hartmann}},
  \ and\ \bibinfo {author} {\bibfnamefont {A.~P.}\ \bibnamefont {Young}},\
  }\href@noop {} {\bibfield  {journal} {\bibinfo  {journal} {Phys. Rev. E}\
  }\textbf {\bibinfo {volume} {102}},\ \bibinfo {pages} {012131} (\bibinfo
  {year} {2020})}\BibitemShut {NoStop}%
\bibitem [{\citenamefont {Charbonneau}\ and\ \citenamefont
  {Tarzia}(2021)}]{charbonneau2021solution}%
  \BibitemOpen
  \bibfield  {author} {\bibinfo {author} {\bibfnamefont {P.}~\bibnamefont
  {Charbonneau}}\ and\ \bibinfo {author} {\bibfnamefont {M.}~\bibnamefont
  {Tarzia}},\ }\href@noop {} {\bibfield  {journal} {\bibinfo  {journal} {J.
  Chem. Phys.}\ }\textbf {\bibinfo {volume} {155}},\ \bibinfo {pages} {024501}
  (\bibinfo {year} {2021})}\BibitemShut {NoStop}%
\bibitem [{\citenamefont {Swendsen}\ and\ \citenamefont
  {Wang}(1986)}]{swendsen1986replica}%
  \BibitemOpen
  \bibfield  {author} {\bibinfo {author} {\bibfnamefont {R.~H.}\ \bibnamefont
  {Swendsen}}\ and\ \bibinfo {author} {\bibfnamefont {J.-S.}\ \bibnamefont
  {Wang}},\ }\href@noop {} {\bibfield  {journal} {\bibinfo  {journal} {Phys.
  Rev. Lett.}\ }\textbf {\bibinfo {volume} {57}},\ \bibinfo {pages} {2607}
  (\bibinfo {year} {1986})}\BibitemShut {NoStop}%
\bibitem [{\citenamefont {Hukushima}\ and\ \citenamefont
  {Nemoto}(1996)}]{hukushima1996exchange}%
  \BibitemOpen
  \bibfield  {author} {\bibinfo {author} {\bibfnamefont {K.}~\bibnamefont
  {Hukushima}}\ and\ \bibinfo {author} {\bibfnamefont {K.}~\bibnamefont
  {Nemoto}},\ }\href@noop {} {\bibfield  {journal} {\bibinfo  {journal} {J.
  Phys. Soc. Japan}\ }\textbf {\bibinfo {volume} {65}},\ \bibinfo {pages}
  {1604} (\bibinfo {year} {1996})}\BibitemShut {NoStop}%
\bibitem [{\citenamefont {Marinari}\ and\ \citenamefont
  {Parisi}(1992)}]{marinari1992simulated}%
  \BibitemOpen
  \bibfield  {author} {\bibinfo {author} {\bibfnamefont {E.}~\bibnamefont
  {Marinari}}\ and\ \bibinfo {author} {\bibfnamefont {G.}~\bibnamefont
  {Parisi}},\ }\href@noop {} {\bibfield  {journal} {\bibinfo  {journal} {EPL}\
  }\textbf {\bibinfo {volume} {19}},\ \bibinfo {pages} {451} (\bibinfo {year}
  {1992})}\BibitemShut {NoStop}%
\bibitem [{\citenamefont {Houdayer}(2001)}]{houdayer2001cluster}%
  \BibitemOpen
  \bibfield  {author} {\bibinfo {author} {\bibfnamefont {J.}~\bibnamefont
  {Houdayer}},\ }\href@noop {} {\bibfield  {journal} {\bibinfo  {journal} {Eur.
  Phys. J. B}\ }\textbf {\bibinfo {volume} {22}},\ \bibinfo {pages} {479}
  (\bibinfo {year} {2001})}\BibitemShut {NoStop}%
\bibitem [{\citenamefont {Kandel}, \citenamefont {Ben-Av},\ and\ \citenamefont
  {Domany}(1990)}]{kandel1990cluster}%
  \BibitemOpen
  \bibfield  {author} {\bibinfo {author} {\bibfnamefont {D.}~\bibnamefont
  {Kandel}}, \bibinfo {author} {\bibfnamefont {R.}~\bibnamefont {Ben-Av}}, \
  and\ \bibinfo {author} {\bibfnamefont {E.}~\bibnamefont {Domany}},\
  }\href@noop {} {\bibfield  {journal} {\bibinfo  {journal} {Phys. Rev. Lett.}\
  }\textbf {\bibinfo {volume} {65}},\ \bibinfo {pages} {941} (\bibinfo {year}
  {1990})}\BibitemShut {NoStop}%
\bibitem [{\citenamefont {Kandel}, \citenamefont {Ben-Av},\ and\ \citenamefont
  {Domany}(1992)}]{kandel1992cluster}%
  \BibitemOpen
  \bibfield  {author} {\bibinfo {author} {\bibfnamefont {D.}~\bibnamefont
  {Kandel}}, \bibinfo {author} {\bibfnamefont {R.}~\bibnamefont {Ben-Av}}, \
  and\ \bibinfo {author} {\bibfnamefont {E.}~\bibnamefont {Domany}},\
  }\href@noop {} {\bibfield  {journal} {\bibinfo  {journal} {Phys. Rev. B}\
  }\textbf {\bibinfo {volume} {45}},\ \bibinfo {pages} {4700} (\bibinfo {year}
  {1992})}\BibitemShut {NoStop}%
\bibitem [{\citenamefont {Cataudella}\ \emph {et~al.}(1994)\citenamefont
  {Cataudella}, \citenamefont {Franzese}, \citenamefont {Nicodemi},
  \citenamefont {Scala},\ and\ \citenamefont
  {Coniglio}}]{cataudella1994critical}%
  \BibitemOpen
  \bibfield  {author} {\bibinfo {author} {\bibfnamefont {V.}~\bibnamefont
  {Cataudella}}, \bibinfo {author} {\bibfnamefont {G.}~\bibnamefont
  {Franzese}}, \bibinfo {author} {\bibfnamefont {M.}~\bibnamefont {Nicodemi}},
  \bibinfo {author} {\bibfnamefont {A.}~\bibnamefont {Scala}}, \ and\ \bibinfo
  {author} {\bibfnamefont {A.}~\bibnamefont {Coniglio}},\ }\href@noop {}
  {\bibfield  {journal} {\bibinfo  {journal} {Phys. Rev. Lett.}\ }\textbf
  {\bibinfo {volume} {72}},\ \bibinfo {pages} {1541} (\bibinfo {year}
  {1994})}\BibitemShut {NoStop}%
\bibitem [{\citenamefont {Cataudella}\ \emph {et~al.}(1996)\citenamefont
  {Cataudella}, \citenamefont {Franzese}, \citenamefont {Nicodemi},
  \citenamefont {Scala},\ and\ \citenamefont
  {Coniglio}}]{cataudella1996percolation}%
  \BibitemOpen
  \bibfield  {author} {\bibinfo {author} {\bibfnamefont {V.}~\bibnamefont
  {Cataudella}}, \bibinfo {author} {\bibfnamefont {G.}~\bibnamefont
  {Franzese}}, \bibinfo {author} {\bibfnamefont {M.}~\bibnamefont {Nicodemi}},
  \bibinfo {author} {\bibfnamefont {A.}~\bibnamefont {Scala}}, \ and\ \bibinfo
  {author} {\bibfnamefont {A.}~\bibnamefont {Coniglio}},\ }\href@noop {}
  {\bibfield  {journal} {\bibinfo  {journal} {Phys. Rev. E}\ }\textbf {\bibinfo
  {volume} {54}},\ \bibinfo {pages} {175} (\bibinfo {year} {1996})}\BibitemShut
  {NoStop}%
\bibitem [{\citenamefont {Coddington}\ and\ \citenamefont
  {Han}(1994)}]{coddington1994generalized}%
  \BibitemOpen
  \bibfield  {author} {\bibinfo {author} {\bibfnamefont {P.~D.}\ \bibnamefont
  {Coddington}}\ and\ \bibinfo {author} {\bibfnamefont {L.}~\bibnamefont
  {Han}},\ }\href@noop {} {\bibfield  {journal} {\bibinfo  {journal} {Phys.
  Rev. B}\ }\textbf {\bibinfo {volume} {50}},\ \bibinfo {pages} {3058}
  (\bibinfo {year} {1994})}\BibitemShut {NoStop}%
\bibitem [{\citenamefont {Prokof'ev}\ and\ \citenamefont
  {Svistunov}(2001)}]{prokof2001worm}%
  \BibitemOpen
  \bibfield  {author} {\bibinfo {author} {\bibfnamefont {N.}~\bibnamefont
  {Prokof'ev}}\ and\ \bibinfo {author} {\bibfnamefont {B.}~\bibnamefont
  {Svistunov}},\ }\href@noop {} {\bibfield  {journal} {\bibinfo  {journal}
  {Phys. Rev. Lett.}\ }\textbf {\bibinfo {volume} {87}},\ \bibinfo {pages}
  {160601} (\bibinfo {year} {2001})}\BibitemShut {NoStop}%
\bibitem [{\citenamefont {Hitchcock}, \citenamefont {S{\o}rensen},\ and\
  \citenamefont {Alet}(2004)}]{hitchcock2004dual}%
  \BibitemOpen
  \bibfield  {author} {\bibinfo {author} {\bibfnamefont {P.}~\bibnamefont
  {Hitchcock}}, \bibinfo {author} {\bibfnamefont {E.~S.}\ \bibnamefont
  {S{\o}rensen}}, \ and\ \bibinfo {author} {\bibfnamefont {F.}~\bibnamefont
  {Alet}},\ }\href@noop {} {\bibfield  {journal} {\bibinfo  {journal} {Phys.
  Rev. E}\ }\textbf {\bibinfo {volume} {70}},\ \bibinfo {pages} {016702}
  (\bibinfo {year} {2004})}\BibitemShut {NoStop}%
\bibitem [{\citenamefont {Wang}, \citenamefont {De~Sterck},\ and\ \citenamefont
  {Melko}(2012)}]{wang2012generalized}%
  \BibitemOpen
  \bibfield  {author} {\bibinfo {author} {\bibfnamefont {Y.}~\bibnamefont
  {Wang}}, \bibinfo {author} {\bibfnamefont {H.}~\bibnamefont {De~Sterck}}, \
  and\ \bibinfo {author} {\bibfnamefont {R.~G.}\ \bibnamefont {Melko}},\
  }\href@noop {} {\bibfield  {journal} {\bibinfo  {journal} {Phys. Rev. E}\
  }\textbf {\bibinfo {volume} {85}},\ \bibinfo {pages} {036704} (\bibinfo
  {year} {2012})}\BibitemShut {NoStop}%
\bibitem [{\citenamefont {Wang}(2005)}]{wang2005worm}%
  \BibitemOpen
  \bibfield  {author} {\bibinfo {author} {\bibfnamefont {J.-S.}\ \bibnamefont
  {Wang}},\ }\href@noop {} {\bibfield  {journal} {\bibinfo  {journal} {Phys.
  Rev. E}\ }\textbf {\bibinfo {volume} {72}},\ \bibinfo {pages} {036706}
  (\bibinfo {year} {2005})}\BibitemShut {NoStop}%
\bibitem [{\citenamefont {Wang}\ and\ \citenamefont
  {Landau}(2001)}]{wang2001efficient}%
  \BibitemOpen
  \bibfield  {author} {\bibinfo {author} {\bibfnamefont {F.}~\bibnamefont
  {Wang}}\ and\ \bibinfo {author} {\bibfnamefont {D.~P.}\ \bibnamefont
  {Landau}},\ }\href@noop {} {\bibfield  {journal} {\bibinfo  {journal} {Phys.
  Rev. Lett.}\ }\textbf {\bibinfo {volume} {86}},\ \bibinfo {pages} {2050}
  (\bibinfo {year} {2001})}\BibitemShut {NoStop}%
\bibitem [{\citenamefont {Dayal}\ \emph {et~al.}(2004)\citenamefont {Dayal},
  \citenamefont {Trebst}, \citenamefont {Wessel}, \citenamefont {Wuertz},
  \citenamefont {Troyer}, \citenamefont {Sabhapandit},\ and\ \citenamefont
  {Coppersmith}}]{dayal2004performance}%
  \BibitemOpen
  \bibfield  {author} {\bibinfo {author} {\bibfnamefont {P.}~\bibnamefont
  {Dayal}}, \bibinfo {author} {\bibfnamefont {S.}~\bibnamefont {Trebst}},
  \bibinfo {author} {\bibfnamefont {S.}~\bibnamefont {Wessel}}, \bibinfo
  {author} {\bibfnamefont {D.}~\bibnamefont {Wuertz}}, \bibinfo {author}
  {\bibfnamefont {M.}~\bibnamefont {Troyer}}, \bibinfo {author} {\bibfnamefont
  {S.}~\bibnamefont {Sabhapandit}}, \ and\ \bibinfo {author} {\bibfnamefont
  {S.~N.}\ \bibnamefont {Coppersmith}},\ }\href@noop {} {\bibfield  {journal}
  {\bibinfo  {journal} {Phys. Rev. Lett.}\ }\textbf {\bibinfo {volume} {92}},\
  \bibinfo {pages} {097201} (\bibinfo {year} {2004})}\BibitemShut {NoStop}%
\bibitem [{\citenamefont {Pleimling}\ and\ \citenamefont
  {Henkel}(2001)}]{pleimling2001anisotropic}%
  \BibitemOpen
  \bibfield  {author} {\bibinfo {author} {\bibfnamefont {M.}~\bibnamefont
  {Pleimling}}\ and\ \bibinfo {author} {\bibfnamefont {M.}~\bibnamefont
  {Henkel}},\ }\href@noop {} {\bibfield  {journal} {\bibinfo  {journal} {Phys.
  Rev. Lett.}\ }\textbf {\bibinfo {volume} {87}},\ \bibinfo {pages} {125702}
  (\bibinfo {year} {2001})}\BibitemShut {NoStop}%
\bibitem [{\citenamefont {Zhang}\ and\ \citenamefont
  {Charbonneau}(2010)}]{zhang2010monte}%
  \BibitemOpen
  \bibfield  {author} {\bibinfo {author} {\bibfnamefont {K.}~\bibnamefont
  {Zhang}}\ and\ \bibinfo {author} {\bibfnamefont {P.}~\bibnamefont
  {Charbonneau}},\ }\href@noop {} {\bibfield  {journal} {\bibinfo  {journal}
  {Phys. Rev. Lett.}\ }\textbf {\bibinfo {volume} {104}},\ \bibinfo {pages}
  {195703} (\bibinfo {year} {2010})}\BibitemShut {NoStop}%
\bibitem [{\citenamefont {Derrida}\ and\ \citenamefont
  {Vannimenus}(1980)}]{derrida1980TM}%
  \BibitemOpen
  \bibfield  {author} {\bibinfo {author} {\bibfnamefont {B.}~\bibnamefont
  {Derrida}}\ and\ \bibinfo {author} {\bibfnamefont {J.}~\bibnamefont
  {Vannimenus}},\ }\href@noop {} {\bibfield  {journal} {\bibinfo  {journal} {J.
  Physique Lett.}\ }\textbf {\bibinfo {volume} {41}},\ \bibinfo {pages} {473}
  (\bibinfo {year} {1980})}\BibitemShut {NoStop}%
\bibitem [{SI()}]{SI}%
  \BibitemOpen
  \href@noop {} {}\bibinfo {note} {See Supplementary Material for details of
  the 1D clustering TM, 2D $T_c$ determination by TM, and TM cluster
  algorithm.}\BibitemShut {Stop}%
\bibitem [{\citenamefont {Jan}, \citenamefont {Coniglio},\ and\ \citenamefont
  {Stauffer}(1982)}]{jan1982study}%
  \BibitemOpen
  \bibfield  {author} {\bibinfo {author} {\bibfnamefont {N.}~\bibnamefont
  {Jan}}, \bibinfo {author} {\bibfnamefont {A.}~\bibnamefont {Coniglio}}, \
  and\ \bibinfo {author} {\bibfnamefont {D.}~\bibnamefont {Stauffer}},\
  }\href@noop {} {\bibfield  {journal} {\bibinfo  {journal} {J. Phys. A}\
  }\textbf {\bibinfo {volume} {15}},\ \bibinfo {pages} {L699} (\bibinfo {year}
  {1982})}\BibitemShut {NoStop}%
\bibitem [{\citenamefont {Hu}\ and\ \citenamefont
  {Charbonneau}(2021)}]{hu2021resolving}%
  \BibitemOpen
  \bibfield  {author} {\bibinfo {author} {\bibfnamefont {Y.}~\bibnamefont
  {Hu}}\ and\ \bibinfo {author} {\bibfnamefont {P.}~\bibnamefont
  {Charbonneau}},\ }\href@noop {} {\bibfield  {journal} {\bibinfo  {journal}
  {Phys. Rev. B}\ }\textbf {\bibinfo {volume} {103}},\ \bibinfo {pages}
  {094441} (\bibinfo {year} {2021})}\BibitemShut {NoStop}%
\bibitem [{\citenamefont {Semerjian}, \citenamefont {Tarzia},\ and\
  \citenamefont {Zamponi}(2009)}]{semerjian2009exact}%
  \BibitemOpen
  \bibfield  {author} {\bibinfo {author} {\bibfnamefont {G.}~\bibnamefont
  {Semerjian}}, \bibinfo {author} {\bibfnamefont {M.}~\bibnamefont {Tarzia}}, \
  and\ \bibinfo {author} {\bibfnamefont {F.}~\bibnamefont {Zamponi}},\
  }\href@noop {} {\bibfield  {journal} {\bibinfo  {journal} {Phys. Rev. B}\
  }\textbf {\bibinfo {volume} {80}},\ \bibinfo {pages} {014524} (\bibinfo
  {year} {2009})}\BibitemShut {NoStop}%
\bibitem [{\citenamefont {Zheng}, \citenamefont {Tarzia},\ and\ \citenamefont
  {Charbonneau}(2022)}]{zmc2022}%
  \BibitemOpen
  \bibfield  {author} {\bibinfo {author} {\bibfnamefont {M.}~\bibnamefont
  {Zheng}}, \bibinfo {author} {\bibfnamefont {M.}~\bibnamefont {Tarzia}}, \
  and\ \bibinfo {author} {\bibfnamefont {P.}~\bibnamefont {Charbonneau}},\
  }\href@noop {} {\bibfield  {journal} {\bibinfo  {journal} {in preparation}\ }
  (\bibinfo {year} {2022})}\BibitemShut {NoStop}%
\bibitem [{\citenamefont {Cataudella}(1992)}]{cataudella1992percolation}%
  \BibitemOpen
  \bibfield  {author} {\bibinfo {author} {\bibfnamefont {V.}~\bibnamefont
  {Cataudella}},\ }\href@noop {} {\bibfield  {journal} {\bibinfo  {journal}
  {Physica A}\ }\textbf {\bibinfo {volume} {183}},\ \bibinfo {pages} {249}
  (\bibinfo {year} {1992})}\BibitemShut {NoStop}%
\bibitem [{\citenamefont {Baillie}\ and\ \citenamefont
  {Coddington}(1991)}]{baillie1991comparison}%
  \BibitemOpen
  \bibfield  {author} {\bibinfo {author} {\bibfnamefont {C.~F.}\ \bibnamefont
  {Baillie}}\ and\ \bibinfo {author} {\bibfnamefont {P.~D.}\ \bibnamefont
  {Coddington}},\ }\href@noop {} {\bibfield  {journal} {\bibinfo  {journal}
  {Phys. Rev. B}\ }\textbf {\bibinfo {volume} {43}},\ \bibinfo {pages} {10617}
  (\bibinfo {year} {1991})}\BibitemShut {NoStop}%
\bibitem [{\citenamefont {Ediger}(2000)}]{ediger2000spatially}%
  \BibitemOpen
  \bibfield  {author} {\bibinfo {author} {\bibfnamefont {M.~D.}\ \bibnamefont
  {Ediger}},\ }\href@noop {} {\bibfield  {journal} {\bibinfo  {journal}
  {Annu.~Rev.~Phys.~Chem.}\ }\textbf {\bibinfo {volume} {51}},\ \bibinfo
  {pages} {99} (\bibinfo {year} {2000})}\BibitemShut {NoStop}%
\bibitem [{\citenamefont {Bouchaud}(2008)}]{bouchaud2008anomalous}%
  \BibitemOpen
  \bibfield  {author} {\bibinfo {author} {\bibfnamefont {J.-P.}\ \bibnamefont
  {Bouchaud}},\ }\href@noop {} {\bibfield  {journal} {\bibinfo  {journal}
  {Anomalous Transport}\ ,\ \bibinfo {pages} {327}} (\bibinfo {year}
  {2008})}\BibitemShut {NoStop}%
\bibitem [{\citenamefont {Tamayo}, \citenamefont {Brower},\ and\ \citenamefont
  {Klein}(1990)}]{tamayo1990single}%
  \BibitemOpen
  \bibfield  {author} {\bibinfo {author} {\bibfnamefont {P.}~\bibnamefont
  {Tamayo}}, \bibinfo {author} {\bibfnamefont {R.~C.}\ \bibnamefont {Brower}},
  \ and\ \bibinfo {author} {\bibfnamefont {W.}~\bibnamefont {Klein}},\
  }\href@noop {} {\bibfield  {journal} {\bibinfo  {journal} {J. Stat. Phys.}\
  }\textbf {\bibinfo {volume} {58}},\ \bibinfo {pages} {1083} (\bibinfo {year}
  {1990})}\BibitemShut {NoStop}%
\bibitem [{SAL()}]{SALRdata}%
  \BibitemOpen
  \href@noop {} {\enquote {\bibinfo {title} {Duke digital repository},}\
  }\bibinfo {howpublished}
  {\url{https://doi.org/10.7924/xxxxxxxxx}}\BibitemShut {NoStop}%
\end{thebibliography}%

\end{document}


\title{Supplementary Material for ``Avoiding critical slowdown in models with SALR interactions''}
\author{Mingyuan Zheng}
\affiliation{Department of Chemistry, Duke University, Durham, North Carolina 27708, United States}
\author{Marco Tarzia}
\affiliation{LPTMC, CNRS-UMR 7600, Sorbonne Universit\'e, 4 Place Jussieu, F-75005 Paris, France}
\affiliation{Institut Universitaire de France, 1 rue Descartes, 75231 Paris Cedex 05, France}
\author{Patrick Charbonneau}
\email[Author to whom correspondence should be addressed. Electronic mail: ]{patrick.charbonneau@duke.edu}
\affiliation{Department of Chemistry, Duke University, Durham, North Carolina 27708, United States}
\affiliation{Department of Physics, Duke University, Durham, North Carolina 27708, United States}
\date{\today}

\maketitle
\renewcommand{\thefigure}{S\arabic{figure}}

\renewcommand{\theequation}{S\arabic{equation}}

\section{1D clustering with Transfer Matrix method}
In the main text, the connection probability of two sites a distance $r$ apart in a 1D chain is given as $\langle \gamma_{i,i+r} \rangle = \mathrm{Tr}[M^r T^{N-r}]/Z$. This expression is exactly equivalent to the spin-spin correlation for a simple Ising model with nearest-neighbor interactions, but approximate when next-nearest-neighbor interactions are also present. An additional contribution is then required to account for sites out of the range $i$ to $i+r$ that get involved in the clustering through next-nearest-neighbor bonds.

\begin{figure}[t]
\centering
\includegraphics[scale=0.6,trim={0.2cm 0.4cm 0 0},clip]{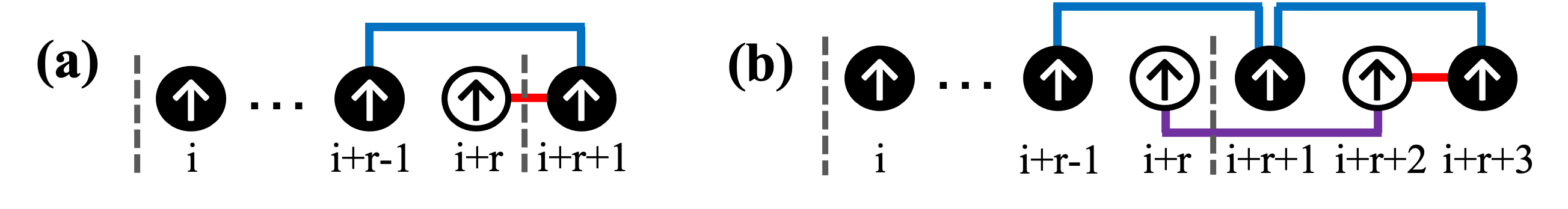}
\caption{Examples of correction cases for $\langle \gamma_{i,i+r} \rangle$ in models with next-nearest-neighbor interactions. The dashed line denotes sites between $i$ and $i+r$. In (a), one site outside gets included in the cluster, and in (b) three sites outside are involved. More outside sites are possible but not repeatedly shown here. The arrows and circle colors have the same meaning as in the main text. Site $i+r$ (white) and site $i+r-1$ (black) belong to two different clusters until a nearest-neighbor bond (red) connects these clusters.}
\label{fig:annni_phase}
\end{figure}

To be exact, we give the explicit expression for $\langle \gamma_{i,i+r} \rangle$ that deals with the boundary between the cluster (described by $M$ in Eq.~(4)) and the environment (described by $T$ in Eq.~(3)), as well as including the aforementioned corrections. Let
\begin{equation}
    \langle \gamma_{i,i+r} \rangle = \frac{2}{Z}(t_1 + t_2),
\end{equation}
which consist of two parts, where $t_1$ denotes the connection probability of site $i$ and $i+r$ through in-between sites, and $t_2$ denotes the  correction of the connection probability involving outside sites. (The factor of two accounts for the fact that clustering can include either up or down spins.)

The first term is
\begin{equation}
\label{eq:t1}
\begin{split}
    t_1 = \mathrm{Tr}[M_0 M^{r-1}M_1T^{N-r-1}],
\end{split}
\end{equation}
where $M_0$ and $M_1$ provide the restriction of boundary sites at the two ends of the 1D cluster existing between sites $i$ and $i+r$,

\begin{equation}
\begin{split}
    M_0 = \left( \begin{array}{ccccc}
    p_1e^{\beta J(1-\kappa)} & (1-p_1)(1-p_2)e^{\beta J(1-\kappa)} & 0 & e^{\beta J(1+\kappa)} & 0 \\
    0 & 0 & 0 & 0 & 0 \\
    p_1e^{-\beta J(1-\kappa)} & (1-p_1)e^{-\beta J(1-\kappa)} & 0 & e^{-\beta J(1+\kappa)} & 0 \\
    0 & 0 & 0 & 0 & 0
     \end{array} \right),
\end{split}
\end{equation}

\begin{equation}
\begin{split}
    M_1 = \left( \begin{array}{cccc}
    e^{\beta J(1-\kappa)} & e^{\beta J(1+\kappa)} & 0 & 0 \\
    p_1p_2e^{\beta J(1-\kappa)} & 0 & 0 & 0 \\
    e^{\beta J(1-\kappa)} & e^{\beta J(1+\kappa)} & 0 & 0 \\
    0 & 0 & 0 & 0 \\
    e^{-\beta J(1-\kappa)} & e^{-\beta J(1+\kappa)} & 0 & 0
     \end{array} \right).
\end{split}
\end{equation}

The second term considers the probability that site $i$ and site $i+r$ are not connected through a certain path within the range $i$ to $i+r$, but that the two clusters to which they respectively belong are then connected at some point out of this range. In fact, it includes corrections corresponding to configurations of $(M_0)_{1,2}$ and $(M_1)_{2,1}$, which miss some situations enabling site $i+1$ or site $i+r$ to be occupied,
\begin{equation}
\begin{split}
    t_2 &= \sum\limits_{j=0}\sum\limits_{k=0}\left(\mathrm{Tr}[M_0^{(occ1)} M'^j M_0^{(occ2)} M^{r-1} M_1^{(1)} M'^k M_1^{(2)} M_1^{(3)} T^{x}]\right.\\
    &+ \left.\mathrm{Tr}[M_0^{(un1)} M'^j M_0^{(un2)} M^{r-1} M_1^{(1)} M'^k M_1^{(2)} M_1^{(3)} T^{x}] \right),
\end{split}
\end{equation}
where $x = N-r-j-k-4$. Note that the terms in the summation decay rapidly with increasing $j$ and $k$. In numerical calculations, we use $j,k < 10$ which ensures that the error is less than $10^{-8}$. 

Because the nature of the correction is similar for all missing cases, we take the cluster end (site $i+r$) as an example. In Eq.~\eqref{eq:t1}, the participation of site $i+r$ to the cluster through its connection to leftward sites has already been considered. Then as a correction, we consider that site $i+r$ has no connection to leftward sites, but it is a part of the cluster through its connection to rightward sites, which are connected to the occupied site $i+r-1$ through a next-nearest-neighbor bond. Therefore, the sites between site $i+r$ and the \textit{real} end of the cluster should all be up spins, and be alternatively occupied and unoccupied through next-nearest-neighbor connection before these two  clusters are connected. Once a connection between two nearest neighbors forms, all of these sites belong to the same cluster. The sites beyond the connected nearest neighbors (the \textit{real} end of the cluster) can then be arbitrary. As a result, we define
\begin{itemize}
    \item $M_1^{(1)}$ as the matrix connecting the configuration probability at site $i+r$ and the start of alternatively occupied/unoccupied sites $i+r+1$;
    \item $M'$ as the probability of an added site being in the occupied/unoccupied series (\textit{i.e.} the added site is connected to the leftward next-nearest neighbor, but not to the nearest neighbor);
    \item $M_1^{(2)}$ as the probability of the added site being the end of the occupied/unoccupied series (\textit{i.e.} the added site is connected to the leftward next-nearest neighbor \emph{and} connected to the leftward nearest neighbor);
    \item $M_1^{(3)}$ as the matrix connecting the connection probability of the \textit{real} end of the cluster and the Boltzmann weight of environment configurations.
\end{itemize}
\begin{equation}
\begin{split}
    M' &= (1-p_1)p_2e^{\beta J(1-\kappa)},  \quad\ \,
    M_1^{(1)} = (0,(1-p_1)p_2e^{\beta J(1-\kappa)},0,0,0)^T,\\
    M_1^{(2)} &= p_1p_2e^{\beta J(1-\kappa)}, \qquad\qquad
    M_1^{(3)} = (e^{\beta J(1-\kappa)},e^{\beta J(1+\kappa)},0,0),
\end{split}
\end{equation}
The summation of probabilities should then be done for different lengths of the occupied/unoccupied series.

Similarly, at the beginning of the cluster, the status of site $i+1$ with respect to the cluster is also affected by its connection to leftwards sites (\textit{i.e.} sites out of the range $i$ to $i+r$), through alternatively occupied/unoccupied site series. Here, we respectively consider the cases of occupied or unoccupied site $i+1$, with connecting matrix $M_0^\mathrm{(occ1)}$ (or $M_0^\mathrm{(un1)}$) and $M_0^\mathrm{(occ2)}$ (or $M_0^\mathrm{(un2)}$), and both make contributions to $\langle \gamma_{i,i+r} \rangle$. The relevant matrices are 
\begin{equation}
\begin{split}
    M_0^\mathrm{(occ1)} &= (p_1e^{\beta J(1-\kappa)}, 0, p_1e^{-\beta J(1-\kappa)}, 0)^T, \quad
    M_0^\mathrm{(un1)} = ((1-p_1)(1-p_2)e^{\beta J(1-\kappa)}, 0, (1-p_1)e^{-\beta J(1-\kappa)}, 0)^T,\\
    M_0^\mathrm{(occ2)} &= ((1-p_1)p_2e^{\beta J(1-\kappa)}, 0, 0, 0, 0),
     \qquad  M_0^\mathrm{(un2)} = (0, (1-p_1)p_2e^{\beta J(1-\kappa)}, 0, 0, 0).
\end{split}
\end{equation}

\section{2D Critical Temperature from Transfer Matrix Method}
Studying algorithmic performances at the critical temperature $T_c$ on a 2D square lattice requires a high-accuracy in determination of $T_c$. We have here adopted the numerical  TM approach of Hu and Charbonneau \cite{hu2021resolving}, which considers a strip of size $L\times\infty$ and leverages the various systems symmetries to reduce computational and memory complexity, and therefore maximally broadens the accessible range, $L\in [12,28]$, to extrapolate results to the thermodynamic $1/L\rightarrow0$ limit. This allows us to numerically estimate the thermodynamic properties of thermodynamic systems with high accuracy. The error on $T_c$ is therefore as low as $10^{-4}$. Even smaller errors could be attained with more efforts, but this precision is sufficient for the system sizes considered in this study. 

\begin{figure}[t]
\centering
\subfigure{
\begin{minipage}{\columnwidth}
\centering
\includegraphics[scale=0.42,trim={0 0 0.2cm 0.2cm},clip]{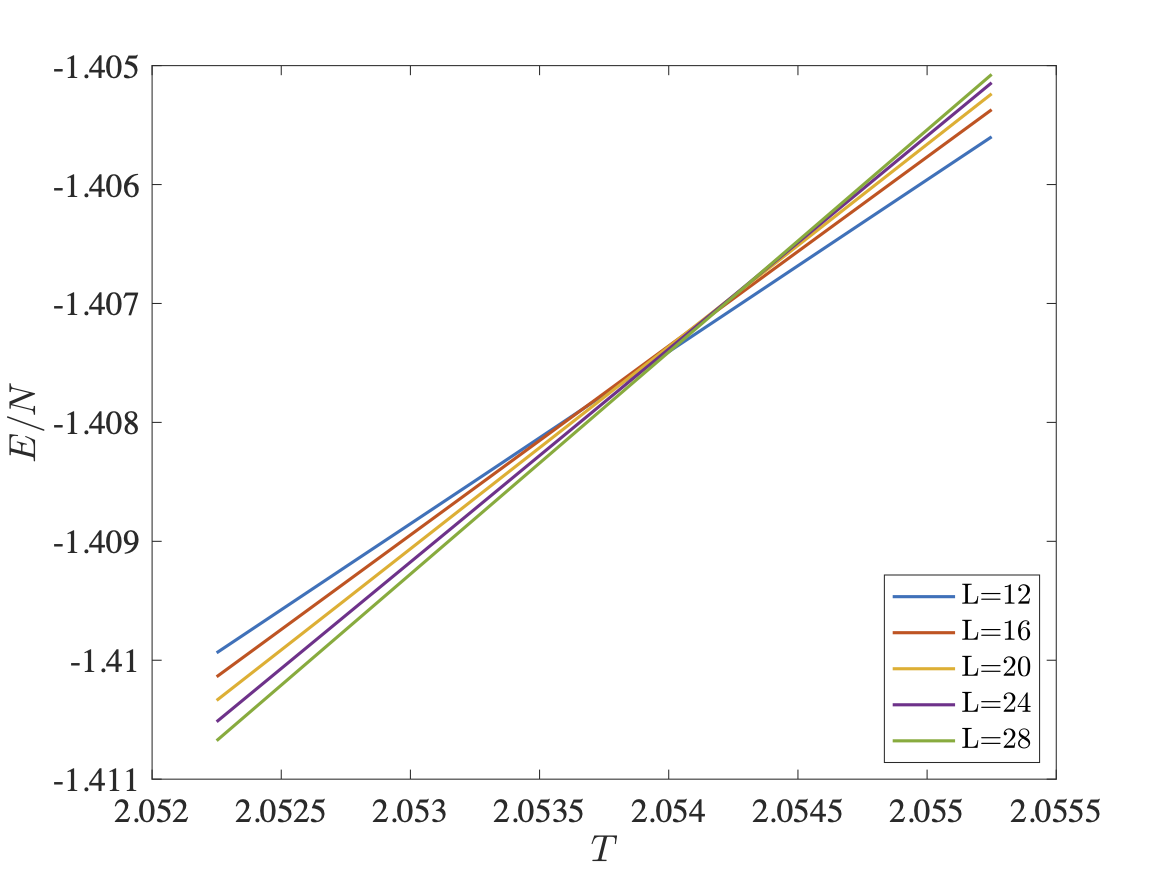}
\includegraphics[scale=0.42,trim={0 0 0.2cm 0.2cm},clip]{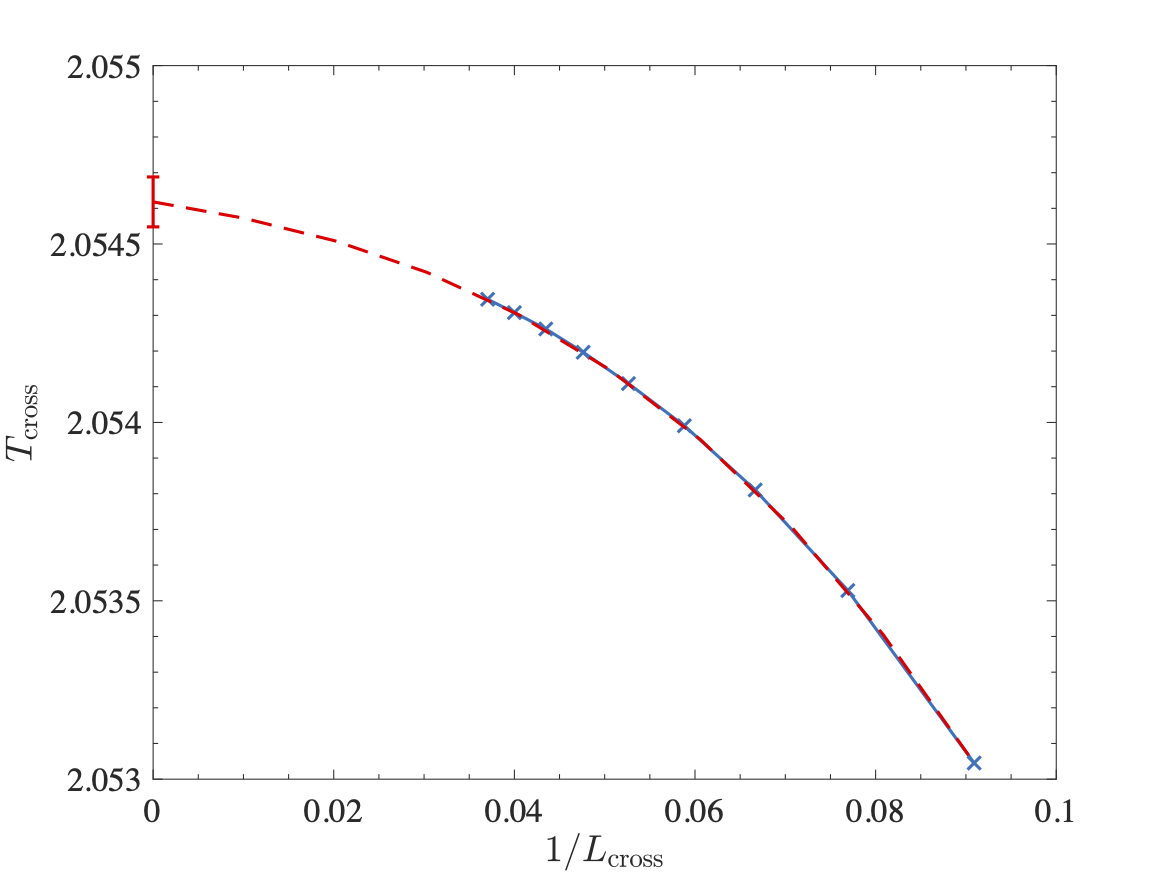}
\end{minipage}
}
\caption{Extrapolation of $T_c$ to $1/L\rightarrow 0$ on a 2D ANNNI model with $\kappa = 0.1$ based on the crossing points in $E-T$ curve. In (b), $T_c = 2.0546(\pm 0.0001)$ is obtained using a cubic form to fit the finite $L$ results.}
\label{fig:annni_scaling}
\end{figure}

We illustrate the approach for $\kappa=0.1$  (Fig.~\ref{fig:annni_scaling}). The system energy $E$ obtained from the first derivative of TM free energy is used as observable. (Observables calculated from second derivatives of the TM free energy, such as the heat capacity, are more sensitive to numerical imprecisions.) As mentioned in Ref.~\cite{hu2021resolving}, $E(T_c)$ is perfectly invariant with $L$ for $\kappa=0$, but varies for $\kappa\not=0$. We therefore identify the crossing point in the $E-T$ curve for every two neighbor system sizes ($L$ is incremented by $2$), and apply the finite-size scaling for $T_{\mathrm{cross}}-1/L_{\mathrm{cross}}$, where $L_{\mathrm{cross}}$ is the arithmetic mean of every two neighbor system sizes. (Using the geometric or the harmonic mean makes little difference.) Extrapolating $T_{\mathrm{cross}}$ to  $1/L_{\mathrm{cross}}\rightarrow 0$ using a cubic form then gives $T_c(\kappa=0.1)=2.0546(1)$.

\section{Transfer Matrix cluster algorithm}
In this section we detail the transfer matrix (TM) cluster algorithm described in the text.
\begin{enumerate}
  \item Randomly select a spin $i$ as the seed site, and check its connection to neighbors in all directions as follows,
  \begin{itemize}
    \item In a selected direction, determine the number of contiguous parallel neighbor sites of the seed site which are not already in the current cluster as $m$.
    \item From Eq.~(8), calculate the associated probability of adding $n < m$ neighbor sites to the current cluster, $s_{n+1}$. If all $m$ sites are added and the growth of clustering is stopped by the interface, its probability is given as $r_{m}$. The probabilities $p^{(n)}$ ($0\leq n \leq m$) are given in Eq.~(9). If this direction is non-axial, then $p_2 = 0$.
    \item With the calculated probabilities, $n \leq m$ neighbor sites in this direction are added to the cluster.
    \item Repeat the above procedure for all directions.
  \end{itemize}
  \item According to the order in which sites are added to the cluster, recursively determine their contiguous neighbor sites being in the same cluster until no more connection is possible, neglecting the direction(s) (both positive and negative) already considered in the previous step.
  \item In order to evaluate the probability of accepting the trial move to flip the constructed cluster, we use the detailed balance criterion
  \begin{equation}
      \prod_{i\in \mathcal{C}_o}p^{(i)} e^{-\beta E_o} \mathrm{acc}(o\rightarrow n) = \prod_{j\in \mathcal{C}_n}p^{(j)} e^{-\beta E_n} \mathrm{acc}(n\rightarrow o),
  \end{equation}
  where $\mathcal{C}_o$ and $\mathcal{C}_n$ are the arrays of the number of added sites in each step constructing the cluster of old configuration and new configuration, respectively, and $E_o$ and $E_n$ denote the energy of old and new configurations, respectively.
  \item Flip the constructed cluster with the calculated acceptance rate.
\end{enumerate}

\bibliography{reference}